\documentstyle[amssymb,12pt]{article}
\begin{document}

\title{Quantum sensitive dependence}
\author{{\it V. I. Man'ko}{\thanks{%
on leave from the P. N. Lebedev Physical Institute, Moscow, Russia}\quad and 
}{\it R.Vilela Mendes}\thanks{%
corresponding author: vilela@cii.fc.ul.pt} \\
%EndAName
{\small Grupo de F\'{i}sica--Matem\'{a}tica,}\\
{\small \ \ Complexo Interdisciplinar, Universidade de Lisboa}\\
{\small \ \ Av. Prof. Gama Pinto, 2, 1699 Lisboa Codex, Portugal}}
\date{}
\maketitle

\begin{abstract}
Wave functions of bounded quantum systems with time-independent potentials,
being almost periodic functions, cannot have time asymptotics as in
classical chaos. However, bounded quantum systems with time-dependent
interactions, as used in quantum control, may have continuous spectrum and
the rate of growth of observables is an issue of both theoretical and
practical concern.

Rates of growth in quantum mechanics are discussed by constructing
quantities with the same physical meaning as those involved in the classical
Lyapunov exponent. A generalized notion of quantum sensitive dependence is
introduced and the mathematical structure of the operator matrix elements
that correspond to different types of growth is characterized.
\end{abstract}



\section{Introduction}

Bounded classical systems that are chaotic, display exponential growth of
initial perturbations and other interesting long-time asymptotics, like
exponential decay of correlations. In contrast, quantum Hamiltonians of
bounded systems with time-independent potentials, having discrete spectrum,
their wave functions are almost periodic functions. For this reason the work
on ``quantum chaos'' has shifted from consideration of long-time properties
to the statistics of energy levels of quantum systems with a chaotic
classical counterpart (for a review of recent work see \cite{Steiner} and
references therein).

However, quantum systems with bounded configuration space but time-dependent
interactions (for example particles in an accelerator subjected to
electromagnetic kicks or the systems used in quantum control) may have
continuous spectrum. Therefore the estimation and control of the rate of
growth, of the perturbed matrix elements of observables, becomes an issue of
both theoretical and practical concern.

In classical mechanics the most important asymptotic indicator of chaotic
behavior is the Lyapunov exponent (an ergodic invariant). Therefore a
natural first step to discuss rates of growth in quantum mechanics seems to
be the construction of a quantum Lyapunov exponent. After a few heuristic
attempts by several authors (see references in \cite{Manko1}) a satisfactory
construction has been achieved \cite{Manko1}, in the sense that the
phase-space observables that are used are exactly the same in classical and
quantum mechanics. It uses the tomographic formulation which describes
conventional quantum mechanics by a set of marginal probability densities 
\cite{Manko2}. Then, the only difference between the classical and the
quantum exponent lies in the time evolution dynamical equation. In Section 2
we recall this result. Translating it from tomographic densities to traces
of operators, it turns out that the quantum Lyapunov exponent measures the
rate of growth of the trace of position and momentum observables starting
from a singular initial density matrix.

A positive Lyapunov exponent corresponds to exponential growth of these
traces. However, the same quantities may serve to characterize other types
of growth, leading to a generalized notion of quantum sensitive dependence.

There are examples where exponential rates of growth (as in classical chaos)
are also found in quantum systems (Sect. 3). However, in many other cases,
quantum mechanics seems to have a definite taming effect on classical chaos.
Therefore, a generalized notion of quantum sensitive dependence,
corresponding to rates of growth milder than exponential, might be of
interest to classify different types of quantum complexity or to
characterize the degree of accuracy achievable in quantum control.

Quantum sensitive dependence is discussed in Sect. 4, as well as the
mathematical structure of the operator matrix elements, in the spectral
representation of the trace, that corresponds to each type of growth. A
convenient unified framework to discuss these matters is the space of
ultradistributions of compact support and their Fourier images \cite{Silva} 
\cite{Pinto}.

\section{The quantum Lyapunov exponent}

As a first step we will rewrite the results of Ref.\cite{Manko1} using
operator traces. In Ref.\cite{Manko1} the quantum Lyapunov exponent along
the phase-space vector $v=\left( v_{1} v_{2}\right) $ is shown to be 
\begin{equation}
\lambda _{v}=\lim_{t\rightarrow \infty }\frac{1}{t}\log \left\| \int
d^{n}X\,d^{n}\mu \,d^{n}\nu e^{iX\bullet {\bf 1}}\left( \left( 
\begin{array}{l}
\nabla _{\mu } \\ 
\nabla _{\nu }
\end{array}
\right) \delta ^{n}(\mu )\delta ^{n}\left( \nu \right) \right) M_{t}\left(
X,\mu ,\nu \right) \,\right\|  \label{2.1}
\end{equation}
where $M_{t}\left( X,\mu ,\nu \right) $%
\begin{equation}
M_{t}\left( X,\mu ,\nu \right) =\int \Pi \left( X,\,\mu ,\,\nu ,\,X^{\prime
},\,\mu ^{\prime },\,\nu ^{\prime },\,t,\,0\right) M_{0}\left( X,\mu ,\nu
\right) dX^{\prime n}d\mu ^{\prime n}d\nu ^{\prime n}  \label{2.2}
\end{equation}
is the time evolved tomographic density \cite{Manko2}, starting from the
initial condition 
\begin{equation}
M_{0}\left( X^{^{\prime }},\mu ^{^{\prime }},\nu ^{^{\prime }}\right)
=\left( \left( v_{1}\circledast \mu ^{\prime }+v_{2}\circledast \nu ^{\prime
}\right) \bullet \nabla _{X^{\prime }}\right) \delta ^{n}\left( X^{\prime
}-\mu ^{\prime }q_{0}-\nu ^{\prime }p_{0}\right)  \label{2.3}
\end{equation}
$a\circledast b$ being defined as 
\[
\left( a\circledast b\right) _{i}=a_{i}b_{i} 
\]

When $\Pi $ is the classical propagator, $\lambda _{v}$, as defined in eq.(%
\ref{2.1}), coincides with the usual classical Lyapunov exponent constructed
from the tangent map. The only difference between classical and quantum
Lyapunov exponents lies in the dynamical law of the propagator, thus
insuring that we are dealing with quantities with the same physical meaning.

For a system with Hamiltonian 
\begin{equation}
H=\frac{{p}^{2}}{2}+V(q)\,,  \label{2.4}
\end{equation}
the evolution equation for the quantum propagator of the tomographic
densities is 
\begin{equation}
\begin{array}{l}
\frac{\partial \Pi }{\partial t}-\mu \bullet \nabla _{\nu }\Pi -\nabla
_{x}V\left( \widetilde{q}\right) \bullet \left( \nu \circledast \nabla
_{X}\Pi \right) \\ 
+\frac{2}{\hbar }\sum_{n=1}^{\infty }(-1)^{n+1}\left( \frac{\hbar }{2}%
\right) ^{2n+1}\frac{\nabla _{i_{1}\cdots i_{2n+1}}V\left( \widetilde{q}%
\right) }{(2n+1)!}\left( \nu \circledast \nabla _{X}\right) _{i_{1}}\cdots
\left( \nu \circledast \nabla _{X}\right) _{i_{2n+1}}\Pi \\ 
=0
\end{array}
\label{2.5}
\end{equation}
with initial condition 
\begin{equation}
\lim_{t\rightarrow t_{0}}\Pi \left( X,\mu ,\nu ,X^{\prime },\mu ^{\prime
},\nu ^{\prime },t,t_{0}\right) =\delta ^{n}\left( X-X^{\prime }\right)
\delta ^{n}\left( \mu -\mu ^{\prime }\right) \delta ^{n}\left( \nu -\nu
^{\prime }\right)  \label{2.6}
\end{equation}
reducing for $\hbar =0$ to the classical evolution equation.

In the tomographic formulation, classical and quantum mechanics are both
described by a set of positive probability distributions $M_{t}\left(
X,\,\mu ,\,\nu \right) $, the $\hbar -$deformation appearing only in the
time-evolution. It is this fact that allows the notion of Lyapunov exponent
to be carried over without ambiguity from classical to quantum mechanics.
However, to relate the Lyapunov exponent to the behavior of operator matrix
elements and the spectral properties of the Hamiltonian, it is more
convenient to rewrite it as a functional of the density matrix $\rho \left(
x,x^{^{\prime }}\right) $. The first step is to consider the Fourier
transform $G_{t}\left( \mu ,\mu \right) $ of the tomographic density $%
M_{t}\left( X,\,\mu ,\,\nu \right) $%
\begin{equation}
G_{t}\left( \mu ,\nu \right) \doteq G_{t}\left( 1,\mu ,\nu \right) =\int
d^{n}X\,e^{iX\bullet {\bf 1}}M_{t}\left( X,\mu ,\nu \right)  \label{2.7}
\end{equation}
and perform the integrals in (\ref{2.1}) to obtain 
\begin{equation}
\lambda _{\left( 
\begin{array}{l}
v_{1} \\ 
v_{2}
\end{array}
\right) }=\lim_{t\rightarrow \infty }\frac{1}{t}\log \left\| 
\begin{array}{l}
\nabla _{\mu }G_{t}\left( \mu ,\nu \right) \mid _{\mu =\nu =0} \\ 
\nabla _{\nu }G_{t}\left( \mu ,\nu \right) \mid _{\mu =\nu =0}
\end{array}
\right\|  \label{2.8}
\end{equation}
Now, using the relation between the tomographic densities and the density
matrix \cite{Manko1}, namely 
\begin{eqnarray}
G_{t}\left( \mu ,\nu \right) &=&\left( \frac{1}{2\pi }\right) ^{n}\int
d^{n}X\,d^{n}pd^{n}xd^{n}x^{^{\prime }}\,e^{i\left( X\bullet {\bf 1-}%
p\bullet \left( x-x^{^{\prime }}\right) \right) }\rho _{t}\left(
x,x^{^{\prime }}\right)  \label{2.9} \\
&&\hspace{3.45cm}\delta ^{n}\left( X-\mu \circledast \left( \frac{%
x+x^{^{\prime }}}{2}\right) +v\circledast p\right)  \nonumber
\end{eqnarray}
one easily obtains 
\begin{equation}
\lambda _{\left( 
\begin{array}{l}
v_{1} \\ 
v_{2}
\end{array}
\right) }=\lim_{t\rightarrow \infty }\frac{1}{t}\log \left\| 
\begin{array}{l}
\textnormal{Tr}\left\{ \rho _{t}x\right\} \\ 
\textnormal{Tr}\left\{ \rho _{t}p\right\}
\end{array}
\right\|  \label{2.10}
\end{equation}
the density matrix at time zero (corresponding to $M_{0}\left( X^{^{\prime
}},\mu ^{^{\prime }},\nu ^{^{\prime }}\right) $ in eq.(\ref{2.3})) being

\begin{equation}
\rho _{0}\left( x,x^{^{\prime }}\right) =-e^{ip_{0}\bullet \left(
x-x^{^{\prime }}\right) }\left\{ \left( v_{1}\bullet \nabla \right) \delta
^{^{n}}\left( q_{0}-\frac{x+x^{^{\prime }}}{2}\right) +iv_{2}\bullet \left(
x-x^{^{\prime }}\right) \delta ^{n}\left( q_{0}-\frac{x+x^{^{\prime }}}{2}%
\right) \right\}  \label{2.11}
\end{equation}
Eq.(\ref{2.10}) means that the quantum Lyapunov exponent measures the
exponential rate of growth of the expectation values of position and
momentum, starting from the initial singular perturbation $\rho _{0}$. This
is a rather appealing and intuitive form for the Lyapunov exponent. That the
quantum Lyapunov exponent should have a form of this type had already been
proposed in Ref.\cite{Vilela1}, based on qualitative physical
considerations. What is not obvious, though, without the tomographic
formulation, is that this is the form that corresponds exactly to the same
physical quantity as the classical Lyapunov exponent. Also non-obvious is
the specific form that the initial singular perturbation $\rho _{0}$ should
take.

Using the time-dependent operators in the Heisenberg picture 
\begin{equation}
\begin{array}{l}
x_{H}\left( t\right) =U^{\dagger }xU \\ 
p_{H}\left( t\right) =U^{\dagger }pU
\end{array}
\label{2.12}
\end{equation}
one has an equivalent form for $\lambda _{\stackrel{\rightarrow }{v}}$%
\begin{equation}
\lambda _{\left( 
\begin{array}{l}
v_{1} \\ 
v_{2}
\end{array}
\right) }=\lim_{t\rightarrow \infty }\frac{1}{t}\log \left\| 
\begin{array}{l}
\textnormal{Tr}^{^{\prime }}\left\{ \rho _{0}x_{H}\left( t\right) \right\} \\ 
\textnormal{Tr}^{^{\prime }}\left\{ \rho _{0}p_{H}\left( t\right) \right\}
\end{array}
\right\|  \label{2.13}
\end{equation}
where we have also defined 
\[
Tr^{^{\prime }}\left\{ \rho _{0}x_{H}\left( t\right) \right\} =Tr\left\{
\rho _{0}x_{H}\left( t\right) \right\} /Tr\left\{ \rho _{0}x_{H}\left(
0\right) \right\} 
\]
Whenever $\rho _{0}x_{H}\left( t\right) $ is a trace class operator, the
term corresponding to Tr$\left\{ \rho _{0}x_{H}\left( 0\right) \right\} $
has no contribution in the $t\rightarrow \infty $ limit. On the other hand,
by taking the appropriate cut-off and a limiting procedure, the above
expression may also make mathematical sense even in some non-trace class
cases.

\section{An example: Kicked motions in the torus}

Let $x_{1},x_{2}\in [-\pi ,\pi )$ be coordinates in the 2-torus $T^{2}$ with
conjugate momenta $p_{1},p_{2}$ and the dynamics be defined by the
Hamiltonian 
\begin{equation}
H=H_{0}+\sum_{n}V\left( x,p\right) \delta \left( t-n\tau \right)  \label{3.1}
\end{equation}
$x\in T^{2}$ , $p\in R^{2}$ and, in particular, $H_{0}=\frac{p^{2}}{2}$ .
Let 
\begin{equation}
{\cal H}={\cal L}^{2}\left( [-\pi ,\pi ),d^{2}x\right)  \label{3.2}
\end{equation}
Physical observables should be self-adjoint operators. Therefore the domain $%
D\left( x_{i}\right) $ of $x_{i}$ is 
\begin{equation}
D\left( x_{i}\right) =\left\{ f\in {\cal H}\right\}  \label{3.3}
\end{equation}
and the domain of $p_{i}$

\begin{equation}
D\left( p_{i}\right) =\left\{ f\in {\cal H}\mid f\left( x_{i}=-\pi \right)
=f\left( x_{i}=\pi \right) \right\}  \label{3.4}
\end{equation}
A convenient basis of vectors in $D\left( p_{i}\right) $ is

\begin{equation}
{\cal H}=\left\{ \left\langle x|q\right\rangle =\frac{1}{\sqrt{2\pi }}%
e^{iq\bullet x}\mid k\in Z^{2}\right\}  \label{3.5}
\end{equation}
The Floquet operator associated to the periodic Hamiltonian $H$ is 
\begin{equation}
U_{F}=U_{0}U_{K}  \label{3.6}
\end{equation}
with $U_{0}=\exp \left( iH_{0}\tau \right) $ and 
\begin{equation}
U_{K}=\exp \left( iV\left( x,p\right) \right)  \label{3.7}
\end{equation}
We will consider different types of kick potentials. Whenever $V\left(
x,p\right) $ is a function of $x$ or $p$ alone,any differentiable function
will generate an unitary operator $U_{K}$ operating in the basis (\ref{3.5}%
). However for kicks of the electromagnetic type, $V\left( x,p\right) =\frac{%
1}{2}\left( x_{i}p_{i}+p_{i}x_{i}\right) $, because

\begin{eqnarray}
D\left( \frac{1}{2}\left( x_{i}p_{i}+p_{i}x_{i}\right) \right) &=&D\left(
x_{i}p_{i}\right) \cap D\left( p_{i}x_{i}\right)  \label{3.8} \\
&=&\left\{ f\in {\cal H}\mid f\left( x_{i}=-\pi \right) =-f\left( x_{i}=\pi
\right) \right\}  \nonumber
\end{eqnarray}
$\frac{1}{2}\left( x_{i}p_{i}+p_{i}x_{i}\right) $ does not generate a
continuous unitary group in (\ref{3.5}) and only a discrete set of kicks
will be acceptable, namely 
\begin{equation}
U_{K}=\exp \left( \frac{1}{2}\left( x\bullet A\bullet p+p\bullet A\bullet
x\right) \right)  \label{3.9}
\end{equation}
$\exp \left( A\right) $ being a 2x2 matrix with integers entries and
determinant one, the last condition resulting from 
\[
e^{\frac{1}{2}\left( x\bullet A\bullet p+p\bullet A\bullet x\right)
}=e^{x\bullet A\bullet p}e^{\frac{1}{2}\textnormal{Tr}A} 
\]
Using the momentum basis (\ref{3.5}) the initial density matrix $\rho _{0}$
of eq.(\ref{2.11}) may be written 
\begin{equation}
\rho _{0}=-2\left( v_{1}\bullet \frac{\partial }{\partial q_{0}}%
+v_{2}\bullet \frac{\partial }{\partial p_{0}}\right) \sum_{k\in
Z^{2}}\left| p_{0}-k\right\rangle e^{i2k\bullet q_{0}}\left\langle
p_{0}+k\right|  \label{3.10}
\end{equation}
and in a position (generalized) eigenstate basis 
\begin{equation}
\rho _{0}=-4\pi \left( v_{1}\bullet \frac{\partial }{\partial q_{0}}%
+v_{2}\bullet \frac{\partial }{\partial p_{0}}\right)
\int_{T^{2}}d^{2}x\left| q_{0}+x\right\rangle e^{i2x\bullet
p_{0}}\left\langle q_{0}-x\right|  \label{3.11}
\end{equation}
The corresponding traces, needed to compute the Lyapunov exponent, are 
\begin{equation}
\textnormal{Tr}\left\{ \rho _{0}U_{t}^{\dagger }\left( 
\begin{array}{l}
x \\ 
p
\end{array}
\right) U_{t}\right\} =-2\left( v_{1}\bullet \frac{\partial }{\partial q_{0}}%
+v_{2}\bullet \frac{\partial }{\partial p_{0}}\right) \sum_{k\in
Z^{2}}e^{i2k\bullet q_{0}}\left\langle p_{0}+k\right| U_{t}^{\dagger }\left( 
\begin{array}{l}
x \\ 
p
\end{array}
\right) U_{t}\left| p_{0}-k\right\rangle  \label{3.12}
\end{equation}
and 
\begin{eqnarray}
\textnormal{Tr}\left\{ \rho _{0}U_{t}^{\dagger }\left( 
\begin{array}{l}
x \\ 
p
\end{array}
\right) U_{t}\right\} &=&-4\pi \left( v_{1}\bullet \frac{\partial }{\partial
q_{0}}+v_{2}\bullet \frac{\partial }{\partial p_{0}}\right)  \label{3.13} \\
&&\hspace{2cm}{\int }_{T^{2}}d^{2}xe^{i2x\bullet p_{0}}\left\langle
q_{0}-k\right| U_{t}^{\dagger }\left( 
\begin{array}{l}
x \\ 
p
\end{array}
\right) U_{t}\left| q_{0}+x\right\rangle  \nonumber
\end{eqnarray}
Another form, that will be used later on, is a spectral decomposition using
the eigenmodes of the Floquet operator. For discrete spectrum 
\begin{equation}
\textnormal{Tr}\left\{ \rho _{0}U_{t}^{\dagger }\left( 
\begin{array}{l}
x \\ 
p
\end{array}
\right) U_{t}\right\} =\sum_{\mu ,\nu }\left\langle E_{\mu }\right| \rho
_{0}\left| E_{\nu }\right\rangle \left\langle E_{\nu }\right| \left( 
\begin{array}{l}
x \\ 
p
\end{array}
\right) \left| E_{\mu }\right\rangle e^{-i\left( E_{\mu }-E_{\nu }\right) t}
\label{3.14}
\end{equation}
and in general 
\begin{equation}
\textnormal{Tr}\left\{ \rho _{0}U_{t}^{\dagger }\left( 
\begin{array}{l}
x \\ 
p
\end{array}
\right) U_{t}\right\} =\int dE_{\mu }dE_{\nu }\rho _{0}\left( \mu ,\nu
\right) \left( 
\begin{array}{l}
x\left( \nu ,\mu \right) \\ 
p\left( \nu ,\mu \right)
\end{array}
\right) e^{-i\left( E_{\mu }-E_{\nu }\right) t}  \label{3.15}
\end{equation}

Three types of potentials will be considered:

(i) $V\left( x,p\right) =0$

This is just free motion on the torus with $U_{t}=\exp \left( i\frac{p^{2}}{2%
}t\right) $. Then 
\[
U_{t}^{\dagger }\left( 
\begin{array}{l}
x \\ 
p
\end{array}
\right) U_{t}=\left( 
\begin{array}{c}
x+tp \\ 
p
\end{array}
\right) 
\]
and from (\ref{3.12}) it follows that Tr$\left\{ \rho _{0}U_{t}^{\dagger
}\left( 
\begin{array}{l}
x \\ 
p
\end{array}
\right) U_{t}\right\} $ is a constant independent of time, implying $\lambda
_{v}=0$. A similar conclusion would be obtained analyzing the spectral
decomposition because the spectrum of the Floquet being discrete in this
case the right-hand-side of Eq.(\ref{3.14}) is an almost periodic function.

Free motion having an irrelevant effect on the computation of the Lyapunov
exponent, we restrict ourselves, for simplicity, to the resonant case, $\tau
=4\pi m$, $m\in Z$, in the next two examples.\medskip

(ii) $V\left( x,p\right) =\alpha $g$\left( x\right) $ and $\tau =4\pi m$, $%
m\in Z$

From 
\[
\left\langle p_{0}+k\right| U_{t}^{\dagger }\left( 
\begin{array}{l}
x \\ 
p
\end{array}
\right) U_{t}\left| p_{0}-k\right\rangle =\left\langle p_{0}+k\right| \left( 
\begin{array}{l}
x \\ 
p+\frac{t}{\tau }\alpha \nabla g\left( x\right)
\end{array}
\right) \left| p_{0}-k\right\rangle 
\]
it follows that $Tr\left\{ \rho _{0}U_{t}^{\dagger }\left( 
\begin{array}{l}
x \\ 
p
\end{array}
\right) U_{t}\right\} $ grows at most linearly with $t$, implying also $%
\lambda _{v}=0$.

In this case the Floquet operator spectrum is continuous but the kernels $%
x\left( \nu ,\mu \right) $ and $p\left( \nu ,\mu \right) $ being 
\begin{eqnarray*}
x\left( \nu ,\mu \right) &\sim &\delta \left( E_{\nu }-E_{\mu }\right) \\
p\left( \nu ,\mu \right) &\sim &\delta ^{^{\prime }}\left( E_{\nu }-E_{\mu
}\right)
\end{eqnarray*}
we obtain the same conclusion from the spectral representation (\ref{3.15}%
).\medskip

(iii) $V\left( x,p\right) =\frac{1}{2}\left( x\bullet A\bullet p+p\bullet
A\bullet x\right) $, $\tau =4\pi m$, $m\in Z$ and $M=\exp \left( A\right)
=\left( 
\begin{array}{ll}
1 & 1 \\ 
1 & 2
\end{array}
\right) $

Let $t=n\tau $. From 
\[
\left\langle p_{0}+k\right| U_{t}^{\dagger }xU_{t}\left|
p_{0}-k\right\rangle =\left( M^{-1^{T}}\right) ^{n}\left\langle
p_{0}+k\right| x\left| p_{0}-k\right\rangle 
\]
and 
\[
\left\langle p_{0}+k\right| U_{t}^{\dagger }pU_{t}\left|
p_{0}-k\right\rangle =M^{n}\left\langle p_{0}+k\right| p\left|
p_{0}-k\right\rangle 
\]
with 
\[
M^{n}=\left( 
\begin{array}{cr}
\omega ^{-2n+1}+\omega ^{2n-1} & -\omega ^{-2n}+\omega ^{2n} \\ 
-\omega ^{-2n}+\omega ^{2n} & \omega ^{-2n-1}+\omega ^{2n+1}
\end{array}
\right) 
\]
\[
M^{-n}=\left( 
\begin{array}{cr}
\omega ^{2n+1}+\omega ^{-2n-1} & -\omega ^{2n}+\omega ^{-2n} \\ 
-\omega ^{2n}+\omega ^{-2n} & \omega ^{2n-1}+\omega ^{-2n+1}
\end{array}
\right) 
\]
and $\omega =\frac{1}{2}\left( 1+\sqrt{5}\right) $, we conclude working out
Eq.(\ref{3.12}) that in this case there is a non-zero Lyapunov exponent $%
\lambda _{v}=2\log \omega $.

It is instructive to find out how the same result may be obtained from the
spectral representation (\ref{3.15}). In the resonant case ($\tau =4\pi m$)\
the eigenstates of the Floquet operator are \cite{Weigert} 
\[
\left| E_{\alpha }\right\rangle =\frac{1}{\sqrt{2\pi }}\sum_{n=-\infty
}^{\infty }e^{-i\alpha n}\left| M^{n}P\right\rangle 
\]
with eigenvalue $\exp \left( -i\alpha \right) $. From 
\begin{eqnarray*}
\left\langle E_{\mu }\right| U_{t}^{\dagger }pU_{t}\left| E_{\alpha
}\right\rangle &=&M^{n}\left\langle E_{\mu }\right| p\left| E_{\alpha
}\right\rangle \\
&=&e^{-i\left( E_{\alpha }-E_{\mu }\right) n\tau }\left\langle E_{\mu
}\right| p\left| E_{\alpha }\right\rangle
\end{eqnarray*}
and the corresponding equation for $x$, it follows that the kernels $x\left(
\nu ,\mu \right) $ and $p\left( \nu ,\mu \right) $ are linear combinations
of 
\[
\delta \left( E_{\alpha }-E_{\mu }-\frac{i}{\tau }\log \lambda _{k}\right) 
\]
$\lambda _{k}$ , $k=1,2$ being the eigenvalues of the matrix $M$. It is the
complex shift in the argument of the delta that converts the complex
exponentials in the spectral decomposition (\ref{3.15}) into an exponential
growing quantity.

Both in case (ii) and (iii) the Floquet spectrum is absolutely continuous.
Nevertheless the rate of growth of the traces is quite different. These
examples suggest that the critical role is actually played by the analytic
nature of the phase-space operator kernels $x\left( \nu ,\mu \right) $ and $%
p\left( \nu ,\mu \right) $. This will further clarified in Sect. 4.

\section{Sensitive dependence in quantum mechanics}

Let us denote the dynamical variable appearing in Eq.(\ref{2.13}) as

\begin{equation}
\Delta \left( t\right) =\left\| 
\begin{array}{l}
\textnormal{Tr}^{^{\prime }}\left\{ \rho _{0}x_{H}\left( t\right) \right\}  \\ 
\textnormal{Tr}^{^{\prime }}\left\{ \rho _{0}p_{H}\left( t\right) \right\} 
\end{array}
\right\|   \label{4.1}
\end{equation}
In the three examples studied in the preceding section this dynamical
quantity shows no growth in the free motion case, polynomial growth for
space-dependent kicks and exponential growth for the electromagnetic-like
kicks. The second case, as well as the study of the standard map in Ref.\cite
{Manko1}, clearly show the taming effect that quantum mechanics has on
classical chaos. Nevertheless $\Delta \left( t\right) $, as defined in Eq.(%
\ref{4.1}), is the quantum observable that corresponds to the notion of
separation of nearby trajectories in the classical case. Therefore, for
example, a polynomial growth of this observable signals an higher degree of
dynamical complexity that no growth at all. Therefore, in view of the
widespread quantum suppression of exponential growth, it makes sense to
characterize different degrees of quantum dynamical complexity by a more
general notion of sensitive dependence. We define:

{\bf Definition} {\it Quantum dynamics is sensitive-dependent in the support
of }$\rho _{0}${\it \ if for any }$T${\it \ and }$M>0${\it , there is a }$t>T
${\it \ such that }$\left( \Delta \left( t\right) /\Delta \left( 0\right)
\right) >M${\it .}

The above definition allows for rates of growth slower than exponential and
even for oscillations of the ratio $\Delta \left( t\right) /\Delta \left(
0\right) $. It only requires it to be unbounded.

In the examples of the preceding section, free motion is not
sensitive-dependent, whereas the $x-$dependent kicks in 3.(ii) are
polynomial sensitive-dependent and the non-local (electromagnetic) kicks in
3.(iii) are exponential sensitive-dependent.

A precise mathematical characterization of when each type of
sensitive-dependence is to be expected, is possible. This uses the
well-known space of ultradistributions of compact support \cite{Silva} \cite
{Pinto} and the corresponding Fourier image.

Let $X_{n}=\left\{ z:z\in {\Bbb C},\left| z\right| >n\right\} $ and ${\cal B}%
_{n}$ be the Banach space of complex functions analytic in $X_{n}$ and
continuous in $\overline{X_{n}}$ with norm 
\begin{equation}
\left\| \phi \right\| _{n}=\sup_{z\in X_{n}}\left| \phi \left( z\right)
\right|  \label{4.2}
\end{equation}
The space of ultradistributions of compact support ${\cal U}_{c}$ is the
inductive limit of the spaces ${\cal B}_{n}$. Its dual is the space of
entire functions. An important subspace of ${\cal U}_{c}$ is the space of
distributions of compact support ${\cal D}_{c}$, the correspondence being
established by the injective (but not surjective) mapping (the Stieltjes
transform) 
\begin{equation}
Sf\left( z\right) =\frac{1}{2\pi i}\int_{\Gamma }\frac{f\left( \lambda
\right) }{\lambda -z}d\lambda  \label{4.3}
\end{equation}
Whenever $f$ is a distribution of compact support, the ultradistribution $%
Sf\in {\cal U}_{c}$ vanishes at infinity.

On the other hand the Fourier transform establishes a correspondence between
the space of ultradistributions of compact support and the space of
functions of exponential growth.

An entire function $\Psi \left( z\right) $ is said to be of exponential
growth if and only if there are constants $\alpha $ and $\beta $ such that $%
\left| \Psi \left( z\right) \right| \leq \alpha e^{\beta \left| z\right| }$, 
$\forall z\in {\Bbb C}$. The vector space of these functions will be denoted
by ${\cal H}_{e}$. The Fourier transform 
\begin{equation}
\left( {\cal F}\phi \right) \left( x\right) =\int_{\Gamma }e^{ix\lambda
}\phi \left( \lambda \right) d\lambda  \label{4.4}
\end{equation}
with $\phi \in {\cal U}_{c}$ is a bijective linear map of ${\cal U}_{c}$
over ${\cal H}_{e}$.

On the other hand the restriction of ${\cal F}$ to the subspace ${\cal D}%
_{c} $ establishes an isomorphism between ${\cal D}_{c}$ and the subspace of 
${\cal H}_{e}$ consisting of entire functions with polynomially bounded
growth on horizontal strips around the real axis.

Now noticing that, in the energy differences $\left( E_{\mu }-E_{\nu
}\right) $ variable, the integral in Eq.(\ref{3.15}) is the Fourier
transform of the kernels

\begin{equation}
{\cal K}\left( \mu ,\nu \right) =\rho _{0}\left( \mu ,\nu \right) \left( 
\begin{array}{l}
x\left( \nu ,\mu \right)  \\ 
p\left( \nu ,\mu \right) 
\end{array}
\right)   \label{4.5}
\end{equation}
we conclude:

{\bf Proposition} {\it A necessary condition for exponential sensitive
dependence in quantum dynamics is that the kernel }${\cal K}\left( \mu ,\nu
\right) ${\it \ as a function of the energy differences }$E_{\mu }-E_{\nu }$%
{\it \ be a member of }${\cal U}_{c}/{\cal D}_{c}${\it . If the kernel
belongs to }${\cal D}_{c}${\it \ then there is, at most, polynomial growth.}

Cases (ii) and (iii) in Section 3 are examples where the kernels belong in
the first case to ${\cal D}_{c}$ and in the second to ${\cal U}_{c}/{\cal D}%
_{c}$ .

\end{document}